# Frequency-Domain Joint Monitoring of Differential Group Delay and Dependent Loss of Optical Single- and Few-Mode Fiber Channels Based on CAZAC Sequences

Linsheng Fan†, Gao Ye†, Zhongliang Sun, Linguo Cao, Hao Shi, Jianwei Tang, Shunfeng Wang, Hengying Xu, Chenglin Bai, Jian Zhao, Weisheng Hu, and Jinlong Wei

*Abstract*—This study focuses on addressing the signal monitoring challenges encountered by optical fiber communication systems involving complicated multi-dimensional channel effects. It is found that when dynamic interference occurs in multiple communication dimensions such as polarizations and modes, traditional monitoring methods face major problems including high cost of using dedicated equipment, difficulty to distinguish channel impairments from each other, and incapability to adapt to the multi-dimensional cases. To address these issues, this paper presents a frequency-domain joint monitoring scheme based on the constant amplitude zero auto-correlation (CAZAC) sequence. The flat spectral characteristics and superior autocorrelation properties of CAZAC sequences are utilized to precisely capture the information of the multi-dimensional channel, which is modeled as a multi-input multi-output (MIMO) system. By employing the estimated channel frequency response, the differential group delay (DGD) and dependent loss (DL) among different dimensions can be extracted and estimated in-service, regardless of the channel transmission dimensionality. The proposed scheme is experimentally verified in both polarization division multiplexing (PDM) and mode division multiplexing (MDM) systems. Experimental results indicate that in the PDM transmission scenario (the 2×2 MIMO case), the proposed scheme enables a polarization-dependent loss (PDL) estimation error consistently below 0.3 dB and a polarization mode dispersion (PMD) estimation accuracy of approximately 0.3 ps; likely, in a 2-mode MDM transmission system (the 4×4 MIMO case), the estimation errors of mode-dependent loss (MDL) and differential mode group delay (DMGD) remain stable at around 0.3 dB and 0.3 ps, respectively. The proposed solution requires no special equipment and is fully compatible with the existing coherent digital signal processing (DSP) architecture. This design not only eliminates costly hardware requirements but also ensures continuous system operation without interruption, dramatically reducing the complexity and cost of channel monitoring while offering a novel approach of in-service signal performance monitoring of multi-dimensional optical communication systems.

## I. INTRODUCTION

As fiber-optic communication systems keep evolving toward larger transmission capacity driven by artificial intelligence (AI) and cloud applications, multidimensional multiplexing technologies, based on polarization multiplexing (PDM), wavelength division multiplexing (WDM), and space division multiplexing (SDM), and so on, boost channel capacity through parallel transmission dimensions. The SDM has emerged as a pivotal method for surpassing the nonlinear Shannon limit of single-mode optical fibers [1]. However, during optical fiber transmission, multidimensional multiplexed signals are subject to various impairments arising from the coupling of multiple physical effects. These impairments include inter-signal crosstalk, pulse broadening induced by chromatic dispersion (CD), temporal shifts of signals in different dimensions caused by differential group delay (DGD), and energy imbalance resulting from dependent loss (DL) among dimensions of polarizations, modes etc. Multi-dimensional and intertwined impairment mechanisms severely constrain the achievable capacity of the system. Moreover, the dynamic coupling across different dimensions considerably complicates the analysis of channel parameters, resulting in highly complex overall transmission characteristics.

Optical performance monitoring (OPM) technology was proposed to primarily monitor the physical layer channel attributes of optical communication system [2]. OPM enables the acquisition of critical parameters necessary for optimizing

This work was supported by the National Key R&D Program of China (2024YFB29NL00100), the National Talent Program, National Natural Science Foundation of China (U23A20282) and Guangdong Basic and Applied Basic Research Foundation (2024A1515010384). (†*Linshen Fan and Gao Ye contributed equally. Corresponding authors: Jinlong Wei and Jian Zhao.*).

Linshen Fan, Gao Ye, Zhongliang Sun, Linguo Cao, Hao Shi, Jianwei Tang, Shunfeng Wang Weisheng Hu, and Jinlong Wei are with Peng Cheng Laboratory (PCL), Shenzhen 518055, China (e-mail: fanlsh@pcl.ac.cn; yeg01@pcl.ac.cn; sunzhl@pcl.ac.cn; caolog@pcl.ac.cn; shih@pcl.ac.cn; tangjw@pcl.ac.cn; wangshf01@pcl.ac.cn; huwsh@pcl.ac.cn; weijl01@pcl.ac.cn).

Gao Ye, Hao Shi, and Jian Zhao are (also) with South China University of Technology, Guangzhou 510641, China (e-mail: 202310193266@mail.scut.edu.cn; 202310193266@mail.scut.edu.cn; zhaojian@scut.edu.cn).

Hengying Xu and Chenglin Bai are with Liaocheng University, Liaocheng 252000, China (e-mail: xuhengving@lcu.edu.cn; baichenglin@lcu.edu.cn).

Color versions of one or more of the figures in this article are available online at http://ieeexplore.ieee.org

the system performance. In particular, the acquired parameters could assist engineers in reducing budget margins in optical link design and support adaptive compensation for system impairments to sustain network stability. Furthermore, OPM enhances the efficiency of spectrum resource utilization [3]. As optical communication systems progress toward higher capacity and multi-dimensional multiplexing while requiring flexible configuration adjustments, there is an urgent demand for innovative monitoring solutions tailored for these systems. The solutions must be capable of jointly estimating multiple impairment parameters, providing real-time online monitoring, ensuring compatibility with existing digital signal processing (DSP) architectures, and offering dimensional scalability.

In the single-mode fiber transmission systems, OPM has been extensively explored. Regarding polarization mode dispersion (PMD) monitoring, prevailing solutions primarily assess the impairment by quantifying variations in the degree of polarization (DOP) of the received signal [4]. Additionally, some experimental studies have incorporated fractional Fourier transform technology on PMD estimation and monitoring [5]. For monitoring polarization-dependent loss (PDL), one approach assesses the extent of loss by analyzing the signal-to-noise ratio (SNR) across various polarization states [6], whereas another infers the loss by computing the projection center of the signal in Stokes space [7]. Notably, compared to assessing individual physical parameters alone, simultaneously monitoring multiple parameters offers enhanced efficiency and cost-effectiveness. Since CD, PDL, and PMD are linear impairments, one can comprehensively monitor these impairments by analyzing the pulse response curve of the equalization filter [8], [9], [10]. In recent years, intelligent algorithms have increasingly been applied in this field. By training computer models to identify impairments characteristics, joint monitoring of multiple parameters has been accomplished [11], [12]. However, it should be noted that although these machine learning-based methods have demonstrated practical value, their core mechanisms still lack adequate theoretical explanations. Consequently, further optimization of the technical solutions is limited.

In SDM transmission systems employing few-mode, multi-mode, or multi-core fibers, channel impairments such as DGD and DL induce signal distortions across multiple dimensions. Moreover, the dynamic interference among multiple dimensions further complicates the analysis of signal features. For the measurement of differential mode group delay (DMGD), existing technologies fall into two main categories. The first one relies on the time-of-flight principle and measures the difference in signal transmission time [13], [14]. The second one involves interference techniques, including microwave interferometry [15], [16], frequency-modulated continuous wave analysis [17], spatial and spectral resolution imaging [18], [19], and Fourier-domain laser mode-locking technology [20]. However, these traditional methods exhibit clear limitations and necessitate the use of an ultra-wideband oscilloscope to meet the high-precision time measurement requirements. In contrast, DSP technology has shown significant advantages [21], [22]. Some studies have shown that DSP-based estimation can accurately estimate DMGD using only instruments with conventional bandwidth (about hundreds of megahertz) [21], thus significantly reducing equipment costs and better supporting engineering applications. On the other hand, the mode-dependent loss (MDL) can be estimated via the channel transmission matrix, which can be obtained by directly inverting the transfer matrix of the multi-input multi-output (MIMO) equalizer in the receiver DSP or by swapping the inputs and outputs of MIMO equalizer and processing them through an additional MIMO equalizer [23]. Although the method for estimating MDL using an additional MIMO equalizer is more accurate, it requires two matrix operations. Such operations impose a significant computational load during real-time processing, thereby increasing the demands on hardware processing capabilities.

As mentioned above, existing monitoring methods generally rely on specialized equipment (e.g., polarization analyzers and interferometers) or require multiple repeated measurements and offline data processing. These conventional schemes suffer from limited functionality and poor scalability, typically detecting only a single impairment parameter alone. Although some studies have successfully implemented joint monitoring of CD, PMD, and PDL in single-mode fiber systems, effective solutions for simultaneously monitoring multiple parameters in multidimensional transmission scenarios, such as in SDM systems, are still missing.

To address this technical gap, this work proposes a novel frequency-domain monitoring solution that utilizes the constant amplitude zero auto-correlation (CAZAC) sequence as the training sequence (TS). The core superiority of this solution lies in its multi-dimensional adaptability, enabling compatibility with various transmission systems, including single-mode fiber, few-mode fiber, and multi-core fiber. The unique properties of CAZAC sequences, specifically their constant amplitude waveforms, flat spectral distributions, and excellent autocorrelation characteristics, are fully leveraged for precise extraction of the multidimensional channel impulse response matrix. By transforming the time-domain response into frequency-domain features, the proposed scheme can simultaneously monitor CD, DGD (including PMD and DMGD), and DL (including PDL and MDL). The performance of this scheme was experimentally evaluated and validated in both PDM (2×2 MIMO) and MDM (4×4 MIMO) systems. The experimental results highlight three key technical advantages of the scheme: First, the scheme achieves high measurement accuracy with a DL estimation error below 0.3 dB, and a DGD estimation error less than 0.3 ps. Second, it is adaptable to various transmission dimensions, with the potential for theoretically unlimited expansion. Third, in comparison with traditional monitoring techniques, it exhibits strong compatibility with existing DSP by eliminating the need for hardware modifications and high-precision clock synchronization, and is also transparent to modulation formats. Incorporation of the monitoring function into the DSP process enables the scheme to seamlessly integrate online and on-service monitoring with data transmission, streamlining system architecture and significantly reducing operation and maintenance costs, which constitutes a critical advantage for practical SDM system deployment.

This paper is organized as follows: Section II addresses the theoretical fundamentals of the proposed method. Section III conducts experimental evaluation of the proposed scheme in a standard single-mode fiber (SSMF) transmission environment, and the measurements are compared with theoretical predictions. Section IV further extends the experiment to few-mode fiber (FMF) scenarios for multidimensional transmission. Finally, Section V concludes the paper with a comprehensive discussion.

## II. PRINCIPLE OF FREQUENCY-DOMAIN JOINT MONITORING ACROSS MULTI-DIMENSIONAL CHANNEL

### A. Optical Channel Model

In this paper, typical coherent optical communication systems are considered [24], [25]. Regardless of whether dual-polarization single-mode or multimode fiber transmission is used, the fiber can be modeled as a MIMO channel. Assuming linear operations in the transmission process and additive white Gaussian noise (AWGN), the MIMO channel model is depicted in Fig. 1 and can be expressed mathematically as [26]:

$$\mathbf{y} = \mathbf{h}\mathbf{x} + \mathbf{n} \tag{1}$$

where $\mathbf{h}$ represents the time-domain transfer function of the MIMO channel, $\mathbf{x}$ denotes the transmitted signal vector, $\mathbf{n}$ is the AWGN, and $\mathbf{y}$ stands for the received signal vector.

In the fiber modeling, inter-channel interactions introduce random crosstalk. To obtain an accurate channel model, a segmented approach is used, and each segment accounts for deterministic impairments. The frequency-domain transfer function of the fiber can be represented in the segmented form to incorporate these factors:

$$\mathbf{H}(\omega) = H_{CD}(\omega)\prod_{l=1}^{L}\mathbf{U}_l\mathbf{\Lambda}_l\mathbf{V}_l \tag{2}$$

where $H_{CD}(\omega)=e^{j\beta_2\omega^2 z/2}$ is the frequency-domain transfer function of CD, $\beta_2$ is the group velocity dispersion parameter (i.e., the second derivative of the propagation constant), and $z$ is the transmission distance. $\mathbf{U}_l$ and $\mathbf{V}_l$ are unitary matrices derived from the singular value decomposition (SVD) of the $\mathbf{H}(\omega)$ / $H_{CD}(\omega)$ for the $l^{th}$ segment. $\mathbf{\Lambda}_l$ represents the diagonal matrix, which contains the DL and DGD information in the $l^{th}$ segment of the fiber. Let $\mathbf{H}_{SOP}(\omega) = \prod_{l=1}^{L}\mathbf{U}_l\mathbf{\Lambda}_l\mathbf{V}_l$, expressed as the generalized Jones matrix associated with polarization and mode coupling. $\mathbf{H}_{SOP}(\omega)$ is represented by a $m \times m$ matrix as:

$$\mathbf{H}_{SOP} = \begin{bmatrix} H_{1,1}(\omega) & H_{1,2}(\omega) & \cdots & H_{1,m}(\omega) \\ H_{2,1}(\omega) & H_{2,2}(\omega) & \cdots & H_{2,m}(\omega) \\ \cdots & \cdots & \cdots & \cdots \\ H_{m,1}(\omega) & H_{m,2}(\omega) & \cdots & H_{m,m}(\omega) \end{bmatrix} \tag{3}$$

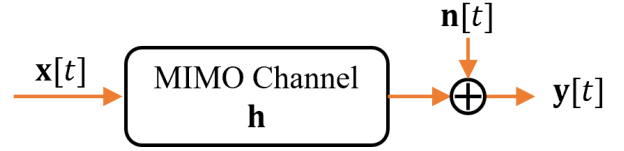

**Fig. 1.** Linear MIMO transmission channel model.

where $m$ is the number of dimensions and represents two polarizations and $m/2$ modes in this paper. The matrix element $H_{p,q}$ denotes the channel frequency response from the $q^{th}$ dimension to the $p^{th}$ dimension, $(p, q = 1, 2, \cdots, m)$. Take SVD for the $\mathbf{H}_{SOP}(\omega)$, we have $\mathbf{H}_{SOP}(\omega) = \mathbf{U}_{ch}(\omega)\mathbf{\Lambda}_{ch}(\omega)\mathbf{V}_{ch}(\omega)$, where the matrix $\mathbf{\Lambda}_{ch}(\omega)$ includes the information of DL and DGD as:

$$\mathbf{\Lambda}_{ch}(\omega) = \text{diag}[\rho_1^{1/2}e^{j\omega\tau_1} \quad \rho_2^{1/2}e^{j\omega\tau_2} \quad \cdots \quad \rho_m^{1/2}e^{j\omega\tau_m}] \tag{4}$$

where $\tau_1, \tau_2, \tau_3, \cdots, \tau_m$ respectively represent the DGD [27], and $\rho_1, \rho_2, \rho_3, \cdots, \rho_m$ contain the DL information. The DL value in dB is expressed by:

$$DL_{dB} = 10\log_{10}\left(\frac{\max(\rho_1, \rho_2, \cdots, \rho_m)}{\min(\rho_1, \rho_2, \cdots, \rho_m)}\right) \tag{5}$$

### B. TS Design and Channel Estimation

The CAZAC sequence $c[n]$ has the properties that its spectrum, $C(\omega_k)$ is flat, that is $|C(\omega_k)|^2 = C$ for all $\omega_k$, and its autocorrelation function has only one non-zero coefficient $ACF[0] = \sum_{n=0}^{N-1}|c[n]|^2 = \sum_{n=0}^{N-1}|C(\omega_k)|^2 / N = C$ [28], [29]. The CAZAC sequence is defined as follows:

$$c[n] = \exp\left\{j\frac{2\pi}{\sqrt{N}}[\text{mod}(n-1,\sqrt{N})+1]\times\left[\left\lceil\frac{n-1}{\sqrt{N}}\right\rceil+1\right]\right\} \tag{6}$$

where $n = 1, 2, \cdots, N$, $N$ is the length of $c[n]$. The CAZAC sequence is essentially a phase shift keying modulated signal and remains fully compatible with payload data modulation. This enables a rate-adaptive transceiver to flexibly switch among various higher-order modulation formats, depending on the required optical SNR (OSNR) and channel capacity. When $c[n]$ cyclically shifts right by $d$ samples, i.e., $c^d[n] = c[n-d]$, the cross-correlation coefficient $CCF_d[n]$ between $c[n]$ and $c^d[n]$ also has only one non-zero value, which can be proved as below:

$$\begin{aligned} CCF_d[n] &= \text{IFFT}\left[C^d(\omega_k)C^*(\omega_k)\right] \\ &= \frac{1}{N}\sum_{k=0}^{N-1}C^d(\omega_k)C^*(\omega_k)e^{j\omega_k n} \\ &= \frac{1}{N}\sum_{k=0}^{N-1}|C(\omega_k)|^2 e^{j\omega_k(n-d)} \\ &= ACF[0]\cdot\delta[n-d] \\ &= \begin{cases} ACF[0], & \text{for } n = d \\ 0. & \text{otherwise} \end{cases} \end{aligned} \tag{7}$$

where $C(\omega_k)$ is the Fast Fourier transform (FFT) of $c[n]$, $\omega_k = 2\pi k/N$. The $\{\cdot\}^*$ represents the conjugate of the matrix.

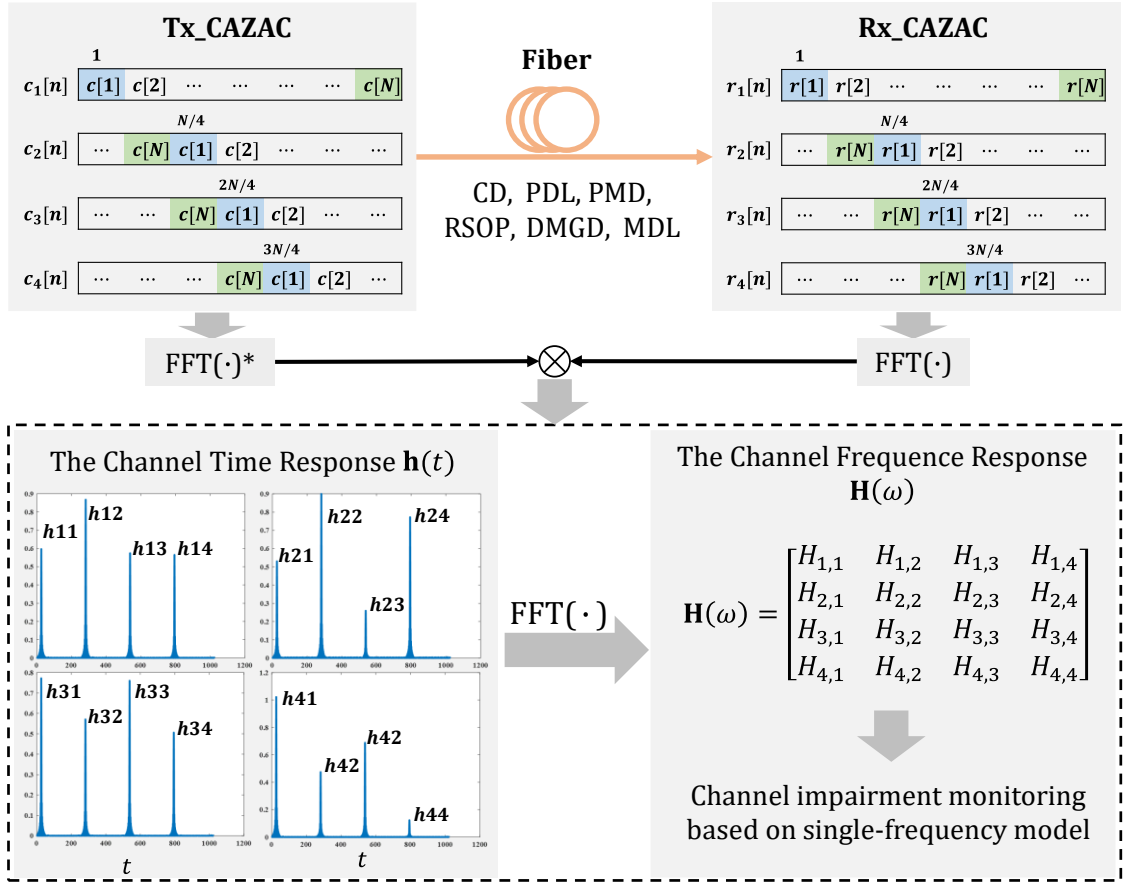

**Fig. 2.** Schematic diagram of the design of TS in a dual-mode transmission system and the corresponding monitoring principles for PDL, DGD, MDL, and DMGD.

The CAZAC sequences can be embedded in the transmission data stream as training data. When the CAZAC sequence is employed as the TS for channel estimation, its length should be determined according to the expected maximum channel impulse response. Since the accuracy of channel estimation is proportional to the TS length and the channel experiences time-varying polarization rotation, periodic insertion of the TS between payload data is required. Based on these properties, in each channel dimension, a CAZAC sequence is transmitted and is cyclically shifted by different increments relative to a reference CAZAC sequence. The length of the reference CAZAC sequence is $N$. To simplify without sacrificing generality, a four-dimensional (two polarizations and two modes) transmission system is considered, and the TS is designed as follows:

$$\begin{aligned}
c_1[n] &= c[n] \\
c_2[n] &= c[n - N/4] \\
c_3[n] &= c[n - 2N/4] \\
c_4[n] &= c[n - 3N/4]
\end{aligned} \quad (8)$$

where $c_1[n]$, $c_2[n]$, $c_3[n]$, and $c_4[n]$ represent the TS in the four dimensions at the transmitter, respectively.

Fig. 2 shows a schematic diagram of the design of TS. All four dimensions simultaneously transmit the CAZAC sequence, with each sequence cyclically shifted by a number of symbols more than the length of channel impulse response. These sequences remain uncorrelated across both time and space. Because the channel impulse response length is unknown beforehand, the simplest approach is to select a fixed interval corresponding to a specific number of symbols, here we set it to be $N/4$.

After propagating through the fiber channel, the transmitted CAZAC sequence experiences various impairments, including CD, DGD and DL. At the receiver, the TSs in the four dimensions are represented as $r_1[n]$, $r_2[n]$, $r_3[n]$, and $r_4[n]$, respectively. Define matrix $\mathbf{R}(\omega_k)$, which has the size of 4×1 with its $p^{\text{th}}$ row as $R_p(\omega_k)$, where $R_p(\omega_k)$ is the spectrum at frequency $\omega_k$ for the received CAZAC sequence in the $p^{\text{th}}$ dimension, $r_p(n)$, $p$=1, 2, 3…, $N$. Similarly, define matrix $\mathbf{C}(\omega_k)$, which also has the size of 4×1 with its $p^{\text{th}}$ row as $C_p(\omega_k)$, where $C_p(\omega_k)$ is the spectrum at frequency $\omega_k$ for the transmitted CAZAC sequence in the $p^{\text{th}}$ dimension, $c_p(n)$. For an arbitrary $\omega_k$, based on their frequency domain relationship, the following operations can be executed:

$$\mathbf{R}(\omega_k) = \mathbf{H}(\omega_k)\mathbf{C}(\omega_k)$$
$$\Downarrow$$
$$\mathbf{R}(\omega_k) \odot \mathbf{C}_1^*(\omega_k) = \mathbf{H}(\omega_k)\mathbf{C}(\omega_k) \odot \mathbf{C}_1^*(\omega_k) \quad (9)$$
$$\Downarrow$$
$$\mathbf{R}'(\omega_k) = \mathbf{H}(\omega_k)\mathbf{C}'(\omega_k)$$

where $\mathbf{H}(\omega_k)$ is the $m \times m$ channel frequency response matrix in (2) at the frequency of $\omega_k$ and its estimation will be described later. The matrix $\mathbf{C}_1^*(\omega_k) = [C_1^*(\omega_k); C_1^*(\omega_k); C_1^*(\omega_k); C_1^*(\omega_k)]$. $\odot$ represents the Hadamard product. In (9), the vector $\mathbf{R}'(\omega_k)$, obtained on the left side with the size of $4 \times 1$, is $[R_1(\omega_k)C_1^*(\omega_k), R_2(\omega_k)C_1^*(\omega_k), R_3(\omega_k)C_1^*(\omega_k), R_4(\omega_k)C_1^*(\omega_k)]$. The vector $\mathbf{C}(\omega_k)$, obtained on the left side with the size of $4 \times 1$, is $[C_1(\omega_k)C_1^*(\omega_k), C_2(\omega_k)C_1^*(\omega_k), C_3(\omega_k)C_1^*(\omega_k), C_4(\omega_k)C_1^*(\omega_k)]$. Then:

$$
\begin{aligned}
R_p(\omega_k)C_1^*(\omega_k) &= \sum_{q=1}^{4} H_{p,q}(\omega_k)C_q(\omega_k)C_1^*(\omega_k) \\
&= \sum_{q=1}^{4} H_{p,q}(\omega_k)C_1(\omega_k)C_1^*(\omega_k)e^{-j\omega_k(q-1)N/4}
\end{aligned}
\tag{10}
$$

where $H_{p,q}(\omega_k)$ is the $p$th-row and $q$th-colum element of $\mathbf{H}(\omega_k)$, $p=1, 2, 3$ and 4. When combined with (7), we have:

$$
\begin{aligned}
\text{IFFT}\left[R_p(\omega_k)C_1^*(\omega_k)\right] &= \sum_{q=1}^{4} \left(h_{p,q}[n] \otimes CCF_{(q-1)N/4}[n]\right) \\
&= \sum_{q=1}^{4} \left(h_{p,q}[n] \otimes (ACF[0] \cdot \delta[n-(q-1)N/4])\right) \\
&= \sum_{q=1}^{4} ACF[0] \cdot h_{p,q}[n-(q-1)N/4]
\end{aligned}
\tag{11}
$$

where $\otimes$ represents the convolution. Therefore, the response of each $h_{p,q}[n]$ can be obtained through time-domain filtering and the time-domain form of the channel response $\mathbf{h}[n]$ can be derived, as shown in Fig. 2. Finally, an FFT is applied to convert $\mathbf{h}[n]$ to the frequency domain, resulting in $\mathbf{H}(\omega_k)$.

### C. Principle of DL Monitoring

From the estimated channel frequency response matrix $\mathbf{H}(\omega)$, frequency responses at multiple frequency points can be extracted to monitor both DGD and DL. Because DGD is frequency-dependent, its estimation is carried out by exploiting differences across various frequencies. In contrast, DL is frequency-independent. Therefore, the channel matrix parameters at a specific frequency are analyzed separately. These impairments manifest distinctly: DL is observed as a power difference between polarizations or modes, whereas DGD appears different temporal shifts of signals in different dimensions. By leveraging these unique properties, decoupling of DL and DGD is achieved. Once the channel matrix $\mathbf{H}(\omega)$ is estimated, simple matrix operations are employed to monitor both DL and DGD.

For the $\mathbf{H}(\omega)$ with dimension $m \times m$, refer to the establishment of the MIMO channel model in Section II.A. The estimation process for DL monitoring is illustrated in Fig. 3(a). At the frequency $\omega_k$, let

$$
\mathbf{W}(\omega_k) = \mathbf{H}^{\mathrm{H}}(\omega_k)\mathbf{H}(\omega_k) \tag{12}
$$

Specifically, $\mathbf{W}(\omega_k)$ eliminates the influence of CD and carrier phase noise. Subsequently, take SVD for $\mathbf{W}(\omega_k)$ and extracted $m$ eigenvalues, i.e., $\lambda_{1,k}, \lambda_{2,k}, \cdots, \lambda_{m,k}$. From the Sections II.A and (12), $\lambda_{i,k} = \rho_{i,k}$, where $i = 1, 2, \cdots, m$. The standard deviation of these eigenvalues or the dB ratio of the square of the largest to the smallest eigenvalue can represent the $DL_{dB}$ [30]. At the frequency point $\omega_k$, the estimates of $DL_{dB}$ in a multi-dimensional transmission scenario can be expressed as:

$$
DL_{dB}(\omega_k) = 10\log_{10}\left(\frac{\max(\lambda_{1,k}, \lambda_{2,k}, \lambda_{3,k}, \cdots, \lambda_{m,k})}{\min(\lambda_{1,k}, \lambda_{2,k}, \lambda_{3,k}, \cdots, \lambda_{m,k})}\right) \tag{13}
$$

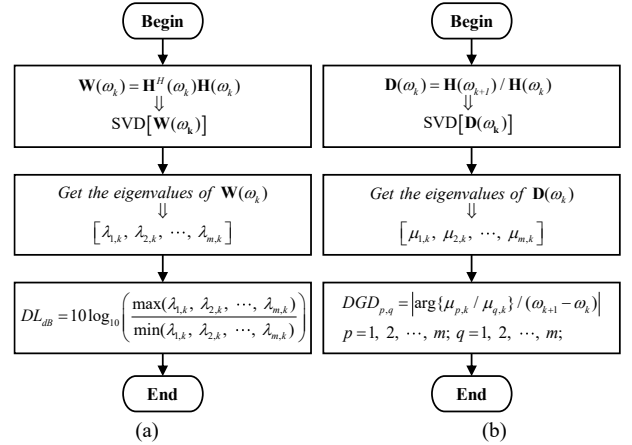

**Fig. 3.** Estimation flowchart. (a) DL estimation process (b) DGD estimation process.

In a two-dimensional transmission scenario, i.e., a single-mode transmission system where $m = 2$, the estimated $DL_{dB}(\omega_k)$ is denoted as, $PDL_{dB}(\omega_k)$ as expressed below:

$$
PDL_{dB}(\omega_k) = 10\log_{10}\left(\frac{\max(\lambda_{1,k}, \lambda_{2,k})}{\min(\lambda_{1,k}, \lambda_{2,k})}\right) \tag{14}
$$

In a four-dimensional transmission scenario, i.e., a two-mode transmission system where $m = 4$, the estimated $DL_{dB}(\omega_k)$ is denoted as $MDL_{dB}(\omega_k)$, as expressed below:

$$
MDL_{dB}(\omega_k) = 10\log_{10}\left(\frac{\max(\lambda_{1,k}, \lambda_{2,k}, \lambda_{3,k}, \lambda_{4,k})}{\min(\lambda_{1,k}, \lambda_{2,k}, \lambda_{3,k}, \lambda_{4,k})}\right) \tag{15}
$$

Both $PDL_{dB}(\omega_k)$ and $MDL_{dB}(\omega_k)$ are monitored at a frequency of $\omega_k$. From Section II.A, it is evident that the value of DL does not depend on frequency. Thereby, PDL and MDL are estimated values remain constant across all frequency points and the final values of their estimated are derived by averaging the estimates across the entire frequency range.

### D. Principle of DGD Monitoring

To estimate DGD, the estimation process is shown in Fig. 3(b). In the channel matrix $\mathbf{H}(\omega_k)$, the DGD manifest as phase differences among the transmission dimensions. However, in the fiber transmission, carrier phase noise and inter-channel interactions also introduce phase variations. Directly estimating DGD from $\mathbf{H}(\omega_k)$ is challenging. Fortunately, the phase variations induced by DGD are frequency-dependent, whereas those caused by carrier phase noise and inter-channel coupling are frequency-independent. This distinction can be utilized to eliminate the effects of carrier phase noise and channel coupling, thereby enabling the estimation of DGD. Therefore, we set $\mathbf{D}(\omega_k) = \mathbf{H}(\omega_{k+1}) / \mathbf{H}(\omega_k)$ to cancel the interference from frequency-independent impairments, and then take SVD for matrix $\mathbf{D}(\omega_k)$. Denoting the eigenvalues of $\mathbf{D}(\omega_k)$ as $\mu_{1,k}$, $\mu_{2,k}$, $\mu_{3,k}$, …, $\mu_{m,k}$, respectively, the DGD can be expressed as:

$$DGD_{p,q} = \left| \arg\{\mu_{p,k} / \mu_{q,k}\} / (\omega_{k+1} - \omega_k) \right| \qquad (16)$$

where $DGD_{p,q}$ represents the DGD between dimensions $p$ and $q$. In a two-dimensional transmission scenario, i.e., a single-mode transmission system where $m = 2$, the estimated $DGD_{p,q}$ is expressed below:

$$PMD = DGD_{2,1} = \left| \arg\{\mu_{2,k} / \mu_{1,k}\} / (\omega_{k+1} - \omega_k) \right| \qquad (17)$$

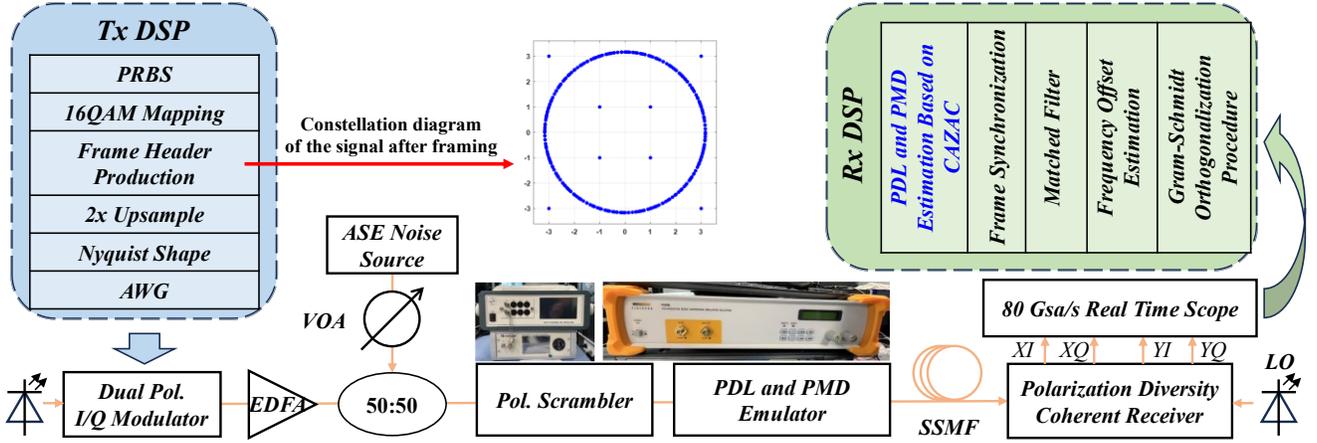

**Fig. 4.** Experimental Setup of Single-Mode Transmission.

## III. Experimental Investigation and Discussion of DL and DGD in Two-Dimensional System

### A. Experimental Setup for Single-Mode Transmission

In this section, a two-dimensional transmission system based on SSMF is established. At the transmitter DSP, a pseudo-random binary sequence (PRBS) is mapped onto a PDM 16 Quadrature Amplitude Modulation (PDM-16QAM) signal. Next, a training sequence, as described in Section II.B, is inserted into the signal to form frames. The resulting 28-Gbaud framed PDM-16QAM signal is then upsampled by a factor of two and filtered using a root-raised cosine (RRC) filter with a roll-off factor of 0.1. Finally, the signal is upsampled again to 128 GSa/s and fed into an arbitrary waveform generator (AWG; Keysight 8199A).

The output of the AWG is modulated by a dual-polarization IQ modulator, whose optical source is a laser with an output power of 15 dBm, wavelength of 1550 nm, and 3-dB linewidth of 100 kHz. The modulated signal is subsequently amplified by an erbium-doped fiber amplifier (EDFA). Thereafter, the EDFA output is combined with amplified spontaneous emission (ASE) noise using a 50:50 coupler. Lastly, the OSNR is controlled by adjusting the ASE noise level via a variable optical attenuator (VOA). A polarization scrambler (EPS1000) is employed in the fiber link to introduce a rate of state of polarization (RSOP) of up to 20 Mrad/s. Following that, a PDL emulator (PDLE-100) and a PMD emulator (PE4200) are incorporated into the system. The PDLE-100 provides a range of 0–20 dB with a resolution of 0.1 dB, whereas the PE4200 introduces DGD in the range of 0–100 ps with a resolution of 0.01 ps. The actual pictures of the EPS1000 and PE4200 are presented in Fig. 4.

Following the introduction of PDL and DGD, the signal propagates through a SSMF toward the receiver. At the receiving end, another laser supplies the local oscillator with an output power of 12 dBm and 3-dB linewidth of 100 kHz. The optical signal is detected using a dual-polarization coherent receiver. Subsequently, a real-time oscilloscope (Lecroy) captures the signal at a sampling rate of 80 GSa/s, thereby converting it into the digital domain for offline processing. Within the receiver DSP, the Gram-Schmidt Orthogonalization Procedure (GSOP) algorithm compensates for IQ imbalance. After frequency offset estimation and compensation, a matched filter processes the signal to maximize its SNR. Finally, frame synchronization is executed and the CAZAC training sequence is extracted to estimate the channel frequency response, which in turn facilitates monitoring of DGD and PDL.

In the two-dimensional transmission experimental system, OSNR is fixed at 22 dB. The preset values of PDL and DGD in the fiber link are adjusted by PDLE-100 and PE4200, respectively. These parameters are monitored at the receiver using the CAZAC TS. The entire channel bandwidth can be utilized to monitor channel impairments. Monitoring accuracy is defined as the difference between the average value calculated from multiple frequency points and the predetermined values for the fiber link, as illustrated below:

$$A_P = \left( \frac{1}{M_f} \sum_{i=1}^{M_f} P_{est,i} \right) - P_{preset} \qquad (18)$$

where $A_P$ denotes the monitoring accuracy for the channel impairment parameter $P_{est}$, while $P_{preset}$ represents the preset magnitude of the channel impairment in the fiber link, and $M_f$ indicates the number of frequency points employed for monitoring within the channel bandwidth. In addition, the standard deviation $\sigma_P$ of the monitoring values obtained from multiple frequency points is used to assess the monitoring stability of the proposed scheme, as expressed below:

$$\sigma_P = \left( \frac{1}{M_f} \sum_{i=1}^{M_f} \left( P_{est,i} - P_{preset} \right) \right) \tag{19}$$

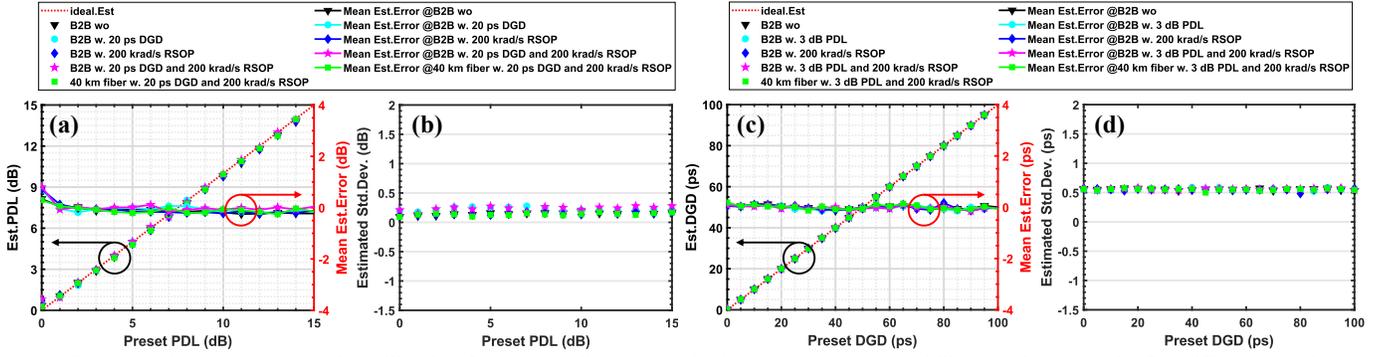

**Fig. 5.** The monitoring accuracy (a) and stability (b) of PDL, as well as the monitoring accuracy (c) and stability (d) of DGD, under five different experimental conditions.

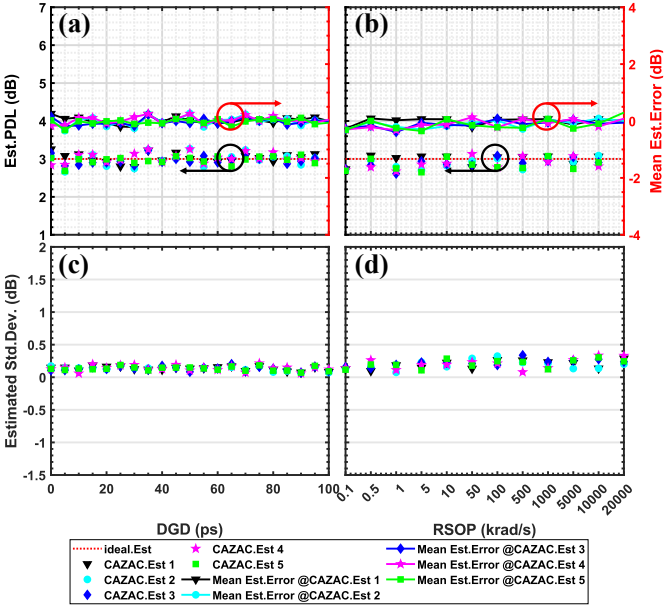

**Fig. 6.** At a fixed preset PDL of 3 dB, the monitoring accuracy (a) and stability (c) of PDL under varying DGD values, as well as the monitoring accuracy (b) and stability (d) of PDL under different RSOP values.

### B. Monitoring of PDL and DGD

In this experiment, the preset PDL values in the fiber link are varied and multiple tests for PDL monitoring are conducted. Initially, the PDL monitoring performance under back-to-back (B2B) transmission is evaluated, with the preset PDL values ranging from 0 to 15 dB in 1-dB increments. To further validate the robustness of the proposed PDL monitoring scheme, its performance is also assessed under additional channel impairments present in the fiber link, including B2B PDL monitoring without any additional impairment, B2B PDL monitoring in the presence of 20-ps

DGD, B2B PDL monitoring under a 200-krad/s RSOP condition, B2B PDL monitoring with both 20-ps DGD and 200-krad/s RSOP, as well as PDL monitoring in a 40-km fiber transmission environment with 20-ps DGD and 200-krad/s RSOP. For each experimental condition, five sets of PDL monitoring data based on the CAZAC sequence are collected to reduce the effects of randomness. In Fig. 5, (a) illustrates the PDL monitoring performance and accuracy metrics, while (b) quantifies the monitoring stability via the standard deviation. Under various channel impairments, the measured PDL values in Fig. 5(a) closely match the preset values, demonstrating the high accuracy and robustness of the proposed PDL monitoring scheme. However, when the preset PDL is 0 dB, a significant deviation in the measured values is observed. Since the PDL monitoring value $PDL_{Est}$ is also influenced by the SNR, its relationship can be expressed as [31]:

$$PDL_{Est} = \frac{PDL^{-1}}{SNR^2} + \frac{2}{SNR} + PDL \tag{20}$$

Therefore, when the PDL is sufficiently small, the impact of SNR on the $PDL_{Est}$ becomes more pronounced, resulting in larger monitoring errors. This represents an inherent limitation of the PDL monitoring scheme based on the eigenvalues of the channel matrix [9], [10], [31], [32]. However, when the preset PDL value exceeds 0 dB, the monitoring error decreases significantly, remaining within 0.2 dB under all channel impairment conditions. Furthermore, for each experimental scenario, the standard deviation of the monitored values obtained from multiple frequency points is computed. As shown in Fig. 5(b), the standard deviations across the five PDL monitoring scenarios are all within 0.3 dB, indicating that the PDL monitoring is stable under these conditions.

In a similar fashion to the PDL monitoring method, each experimental condition produced five sets of DGD monitoring data derived from the CAZAC sequence, and their average are computed to reduce randomness. Fig. 5(c) illustrates the DGD monitoring performance and accuracy metrics, whereas Fig. 5(d) presents the monitoring stability quantified by the standard deviation. In Fig. 5(c), despite various channel impairments, the measured DGD values closely matched the preset values, with errors confined within 0.2 ps, thus confirming the accuracy and robustness of the monitoring approach. Moreover, Fig. 5(d) demonstrates that the standard deviations of the monitoring values across different frequency points for each experimental scenario remained below 0.6 ps, indicating stable DGD monitoring under these conditions.

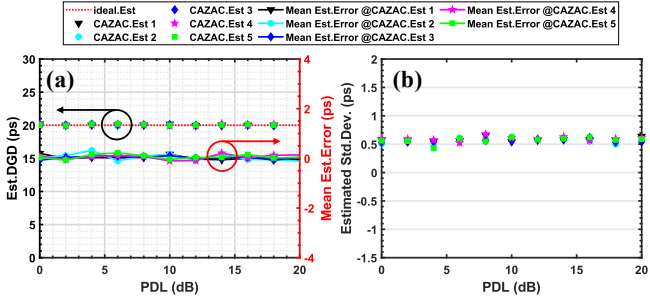

**Fig. 7.** The monitoring accuracy (a) and stability (b) of DGD with a fixed preset value of 20 ps under varying PDL values.

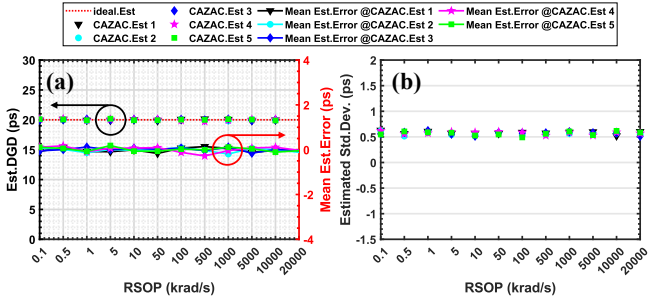

**Fig. 8.** The monitoring accuracy (a) and stability (b) of DGD with a fixed preset value of 20 ps under varying RSOP values.

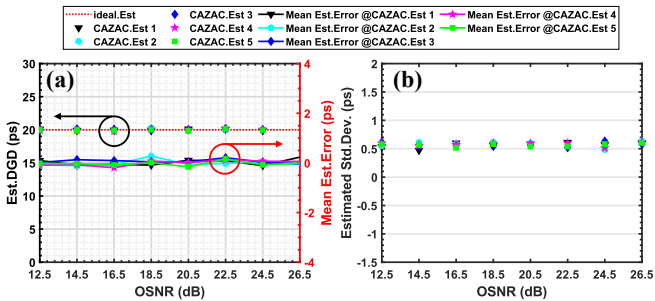

**Fig. 9.** The monitoring accuracy (a) and stability (b) of DGD with a fixed preset value of 20 ps under varying OSNR values.

To further evaluate the reliability of the proposed scheme, five independent experiments are conducted under fixed PDL or DGD conditions while varying other channel impairment. The five sets of collected data are denoted as CAZAC.Est 1, CAZAC.Est 2, CAZAC.Est 3, CAZAC.Est 4, and CAZAC.Est 5, respectively. In Fig. 6(a)-(b), the preset PDL is fixed at 3 dB and the RSOP at 200 krad/s, while the DGD is varied from 0 to 100 ps in 5-ps increments. With the monitoring error consistently below 0.3 dB and the standard deviation of the monitoring values below 0.4 dB, the results show that the estimation performance of PDL is unaffected by DGD. Similarly, in Figs. 6(c)-(d), with the preset PDL fixed at 3 dB and the DGD fixed at 20 ps, the RSOP is varied over the range [0.1, 0.5, 1, 5, 10, 50, 100, 500, 1000, 1500, 10000, 20000] krad/s. The monitoring errors also keep below 0.3 dB and standard deviations under 0.4 dB, which indicate that the PDL estimation can tolerate ultrafast polarization variations.

The performance for DGD monitoring under varying conditions of PDL, RSOP, and OSNR are also shown in Figs. 7-9, with the preset DGD fixed at 20 ps. In Fig. 7, the RSOP is set at 200 krad/s and PDL is varied from 0 to 20 dB in 1-dB increments. In Fig. 8, the PDL is fixed at 3 dB and the RSOP is varied across the range [0.1, 0.5, 1, 5, 10, 50, 100, 500, 1000, 1500, 10000, 20000] krad/s. In Fig. 9, with PDL fixed at 3 dB and RSOP fixed at 200 krad/s, the OSNR is varied from 12.5 to 16.5 dB in 1-dB increment. In all three DGD monitoring scenarios, the monitoring error is kept below 0.3 ps, and the standard deviation of the monitoring values remains below 0.7 ps.

These results indicate that the monitoring performance of PDL and DGD do not affect each other, and exhibit strong robustness against RSOP and ASE noise.

## IV. Experimental Investigation and Discussion of DL and DGD Monitoring in Four-Dimensional System

### A. Experimental Setup for Four-Dimensional Transmission

In this section, a four-dimensional transmission system via two homogeneous modes is established to experimentally validate the proposed joint monitoring scheme for MDL and DMGD, as shown in Fig. 10. At the transmitter, two sets of 31 Gbaud PDM-16QAM signals are generated using DSP. After pulse-shaping filtering, the signals are fed into an AWG and then drive two transmitters that convert the electrical signals into optical signals (as described in Section III). The outputs from the two transmitters are subsequently amplified by EDFAs, and the amplified signals are combined with ASE noise through a 50:50 coupler. Two VOAs are used to adjust the output of the ASE noise source, thereby controlling the OSNR of each channel. The amplified signals are then transmitted through a mode multiplexer (PROTEUS-S, CAILABS) to excite two homogeneous modes HG01 and HG02 on a 20-m FMF. At the receiver, the transmitted signal is decomposed into two orthogonal signals by demultiplexer, which are then amplified by EDFA and received separately by coherent receivers. Finally, four IQ signals are captured by real-time oscilloscope (LeCroy) at an 80-GSa/s sampling rate and converted to the digital domain for offline processing. The receiver DSP follows the same processing flow as described in Section III.A. After frame synchronization, CAZAC TS are extracted from the two transmission modes to estimate the channel frequency response, enabling joint monitoring of MDL and DMGD.

Due to equipment limitations, it is challenging to directly introduce customized MDL and DMGD into the fiber link. Therefore, we implement an alternative approach by introducing these impairments at the receiver end through DSP, as described below:

$$\mathbf{E}_r'[n] = \mathbf{SOP}_{martix}^{-1}\mathbf{MDL}_{martix}\mathbf{DMGD}_{martix}\mathbf{SOP}_{martix}\mathbf{E}_r[n] \quad (21)$$

where $\mathbf{E}_r[n]$ represents the signal collected by the receiver in the digital domain; $\mathbf{MDL}_{matrix}$ = diag[$\rho_1^{1/2}$, $\rho_2^{1/2}$, $\rho_3^{1/2}$, $\rho_4^{1/2}$], which is the matrix to introduce MDL; $\mathbf{DMGD}_{matrix}$ = diag[$e^{j\omega\tau_1}$, $e^{j\omega\tau_2}$, $e^{j\omega\tau_3}$, $e^{j\omega\tau_4}$] is the matrix to induce DMGD; and $\mathbf{E}_r'[n]$ is the signal after both MDL and DMGD have been incorporated. $\rho_1$, $\rho_2$, $\rho_3$ and $\rho_4$ represent the MDL values across the four transmission dimensions. By selecting the minimum value for $\rho_1$ as a reference and adjusting the maximum value within $\rho_2$, $\rho_3$, and $\rho_4$, the target MDL is determined. $\tau_1$, $\tau_2$, $\tau_3$ and $\tau_4$ represent the DGD values across the four transmission dimensions. With the DGD between polarizations set to 0 via $\tau_1 = \tau_2$ and $\tau_3 = \tau_4$, only DMGD is considered; consequently, by using $\tau_1$ and $\tau_2$ as reference dimensions and adjusting $\tau_3$ and $\tau_4$, the target DMGD is realized. It should be noted that random 4×4 matrix is generated, its SVD decomposition results in a unitary matrix $\mathbf{SOP}_{matrix}$, which is used as the polarization coupling matrix in four dimensions.

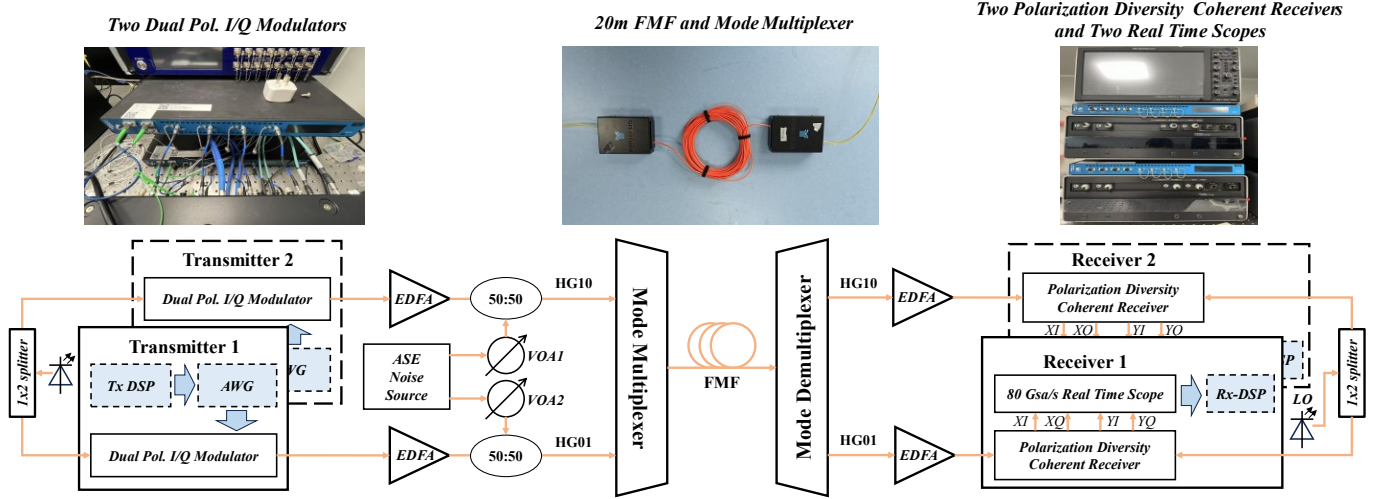

**Fig. 10.** Experimental setup of few-modes transmission.

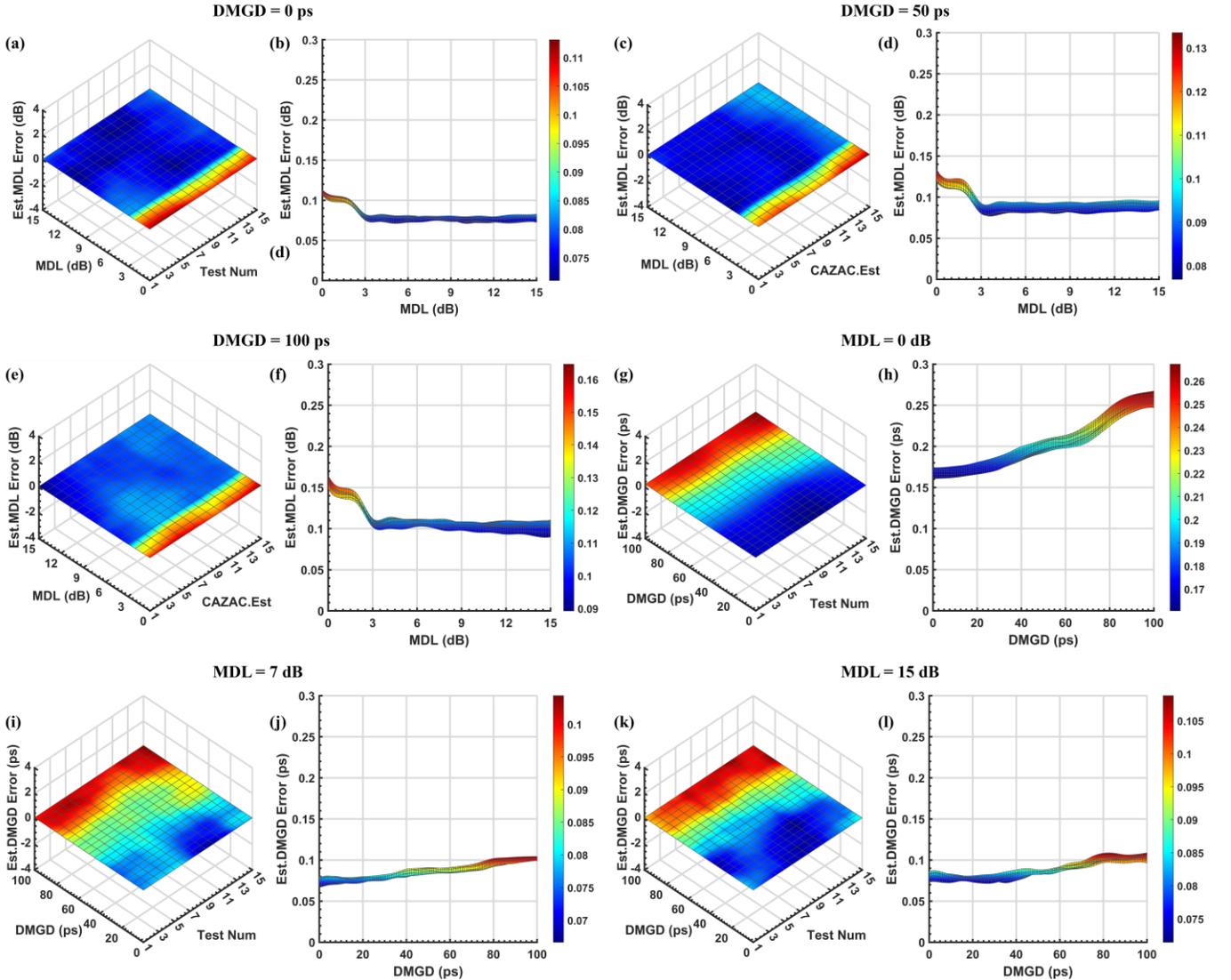

**Fig. 11.** The MDL monitoring results of 15 sets under different DMGD conditions as follows: (a)-(b) DMGD = 0 ps, (c)-(d) DMGD = 50 ps, (e)-(f) DMGD = 100 ps. The DMGD monitoring results of 15 sets under different MDL conditions are shown as: (g)-(h) MDL = 0 dB, (i)-(j) MDL = 7 dB, (k)-(l) MDL = 15 dB. (b), (d), (f), (h), (j), (l) represent the projections of (a), (c), (e), (g), (i), (k) onto the planes formed by the X and Z axes, respectively.

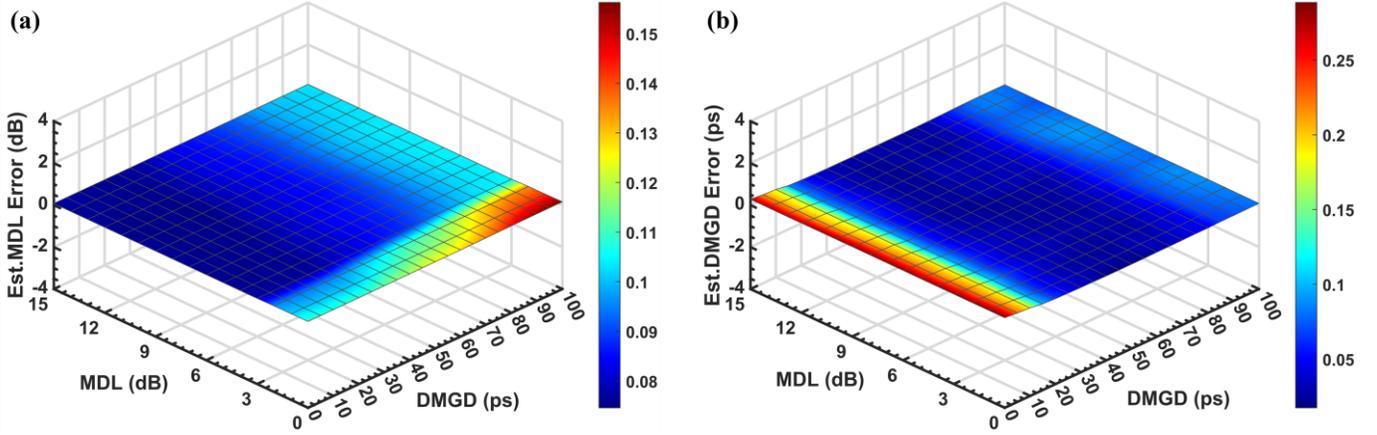

**Fig. 12.** The simultaneous monitoring errors of MDL and DMGD are evaluated over the ranges MDL = [0, 15] dB and DMGD = [0, 100] ps. (a) shows the MDL monitoring error, while (b) presents the DMGD monitoring error.

To eliminate inherent impairments in the fiber link, we adjust the EDFA to neutralize MDL in the link, and tune delay lines to counteract DMGD in the link. The OSNR is maintained at a constant 22 dB.

### B. Monitoring of MDL and DMGD

Initially, MDL is applied over a range of 0 to 15 dB in 1-dB increments. For each preset MDL value, 15 monitoring trials are performed with distinct random seeds, and three cases with additional DMGD values of 0 ps, 50 ps, and 100 ps are evaluated. Fig. 11 (a)-(f) presents the monitoring results. In subfigures (a), (c), and (e), the three-dimensional surfaces represent MDL monitoring errors over MDL within [0,15] dB. The surface remains close to zero with only minimal fluctuations, indicating the high accuracy and robustness of MDL monitoring. Subfigures (b), (d), and (f) display the projections of these surfaces onto a plane defined by MDL and the estimated MDL error. It can be observed that as MDL approaches zero, the estimation error tends to increase. The reason is similar to PDL estimation in the two-dimensional case. Under all test cases, the MDL monitoring error remain below 0.3 dB, confirming that the proposed scheme achieves accurate MDL monitoring and is unaffected by DMGD.

Additionally, DMGD is introduced over a range from 0 to 100 ps in 5-ps increments. For each preset DMGD value, 15 monitoring trials are performed using different random seeds. Three scenarios with additional MDL of 0 dB, 7 dB, and 15 dB are considered, and the corresponding monitoring results are presented in Fig. 11 (g)-(l). In sub-figures (g), (i), and (k), the three-dimensional surfaces representing the DMGD monitoring error and subfigures (h), (j), and (l) display the projections of these surfaces. It is observed that the DMGD monitoring error remains below 0.3 ps in all test scenarios. The results confirm that the proposed scheme achieves accurate DMGD monitoring and is unaffected by MDL.

Finally, the full ranges of MDL (0-15 dB) and DMGD (0-100 ps) are explored, considering all scenarios in which both impairments coexist. The monitoring results are presented in Fig. 12, where subfigure (a) shows the MDL monitoring error and (b) depicts the DMGD monitoring error. Both three-dimensional surfaces appear relatively flat, and the monitoring errors consistently remain below 0.3. These findings further reinforce the high accuracy and stability of the proposed joint monitoring scheme for concurrent MDL and DMGD impairments.

### V. CONCLUSION

This study tackles the technical challenge of concurrent monitoring of channel impairments in multi-dimensional optical fiber communication systems by introducing a frequency-domain processing method based on CAZAC sequences. Exploiting the flat spectral characteristics and excellent autocorrelation properties of CAZAC sequences, the channel frequency response is accurately determined. By leveraging the distinctive frequency characteristics of various impairments, we successfully isolate DGD and DL from other channel impairments. Incorporating the SVD method achieves the joint monitoring of DGD and DL. We experimentally validated our approach in two distinct transmission systems: a two-dimensional system based on SSMF, and a four-dimensional transmission system utilizing few-mode fiber with two homogeneous modes. Across all test scenarios, the estimation error for DL remain below 0.3 dB, while the estimation error for DGD is maintained under 0.3 ps. The experimental results demonstrate the high accuracy and robustness of our proposed method.

From an implementation perspective, the proposed monitoring solution can be integrated seamlessly into existing DSP systems without requiring additional high-bandwidth oscilloscopes, calibration downtime, or hardware modifications. By merging monitoring capabilities with standard communication operations, the design simplifies system architecture while significantly reducing operational and maintenance costs. The applicability of method across various optical communication scenarios paves new technical pathways for developing flexible and intelligent optical networks.